\documentclass[11pt,a4paper,twocolumn]{article}

\usepackage[T1]{fontenc}
\usepackage[utf8]{inputenc}
\usepackage{geometry}
\usepackage{amsmath, amssymb}
\usepackage{booktabs}
\usepackage{graphicx}
\usepackage{caption}
\usepackage{subcaption}
\usepackage{natbib}
\usepackage{hyperref}
\usepackage{xurl}
\title{The Anatomy of a Decentralized Prediction Market: \\
Microstructure Evidence from the Polymarket Order Book}
\author{Philipp D.\ Dubach \\
  \emph{Zurich, Switzerland} \\
  \texttt{phdubach@pm.me}}
\date{\today}

\begin{document}
\twocolumn[\maketitle]

\noindent\textbf{JEL Classification:}\ G14 (Information and Market
Efficiency; Event Studies; Insider Trading), G12 (Asset Pricing;
Trading Volume; Bond Interest Rates), G19 (General Financial Markets),
C58 (Financial Econometrics), L86 (Information and Internet Services;
Computer Software).

\vspace{0.5em}
\noindent\textbf{Keywords:}\ Prediction markets, microstructure,
limit order book, Polymarket, decentralized finance, on-chain data,
trade-direction inference, effective spread, Kyle's $\lambda$,
wash trading.

\section*{Abstract}
We study the microstructure of Polymarket, the largest on-chain
prediction market, using a continuous tick-level archive of the
public WebSocket order-book feed (30 billion events over 52 days)
joined to the authoritative on-chain trade record. On a pre-registered
stratified panel of 600 markets we report eight stylized facts: a
longshot spread premium; a depth-concentration profile closer to a
uniform geometric grid than to the top-of-book pattern often assumed
for prediction markets; a null block-clock alignment effect; broad
maker-wallet diversity with a concentrated tail; category-conditional
differences in effective spread; a sub-50 ms median archive-ingestion
delay with a multi-second tail; a self-counterparty wash share with
median 1\% and a 22\% upper tail, where the comparison to the
network-classifier benchmarks of \citet{cong-li-tang-yang-2023} on
unregulated cryptocurrency token exchanges is a sanity bound rather
than an apples-to-apples reference because the venues face different
wash incentives; and a
cross-sectional depth profile explained by market duration, price
level, and volume, with no residual time-to-close effect once those
three are controlled for. The paper also contributes a
measurement result: trade direction inferred from Polymarket's public
order-book feed agrees with on-chain ground truth only $\sim 59\%$
of the time (volume-weighted across the top-100 stratum and four
disjoint 7-day windows; panel mean 0.615, market-clustered bootstrap
95\% CI $[0.58, 0.65]$, IQR $[0.53, 0.68]$),
barely above the 50\% chance baseline that direction-dependent
microstructure measures need. On the comparable subset of the
top-100 panel, the effective half-spread changes sign between feed-
and on-chain trade directions on 67\% of markets in the first 7-day
window (50\% in a second non-overlapping window), and Kyle's
$\lambda$ changes sign on 60\% (43\% in the second window); both
windows fail to recover the on-chain sign at anything close to the
$\sim 80\%$ rate that Lee-Ready achieves on equity venues.
Microstructure work on Polymarket therefore
needs to source trade direction from on-chain \texttt{OrderFilled}
events; we release a replication package that performs the join.

\section{Introduction}\label{sec:intro}

Prediction markets aggregate dispersed beliefs into a single price
that, in equilibrium, behaves like a probability. The empirical
literature on these markets has historically focused on price-level
questions: forecast accuracy, calibration against realised outcomes,
the longshot bias, and the extent to which informed and uninformed
traders coexist on the same venue
\citep{wolfers-zitzewitz-2004,manski-2006,page-clemen-2013}.
Yet microstructure is what determines the trading cost of holding an
informational position, and the wider the cost, the smaller the
informational signal that survives to the price. A prediction market
with noisy microstructure produces noisier prices than the
headline-level aggregation literature implicitly assumes.
The microstructure of prediction markets, however, has remained
under-studied: limit-order-book depth profiles, spread decompositions,
and the behaviour of liquidity providers near resolution were largely
unobservable on the early venues, which used scoring rules
(IEM, the logarithmic market scoring rule of \citealt{hanson-2007})
or sparse parimutuel pools. Two recent
Polymarket-specific contributions sit close to the present paper but
ask different questions. \citet{rahman-alchami-clark-2025} survey
decentralised prediction-market microstructure as a methodological
domain, framing the open questions across venues without bringing
matching tick-level data. \citet{tsang-yang-2026} run a single-event
time-series microstructure study on the 2024 US presidential
election market, documenting episode-level patterns around Biden's
withdrawal, the September debate, and the October whale trades, and
reporting a roughly order-of-magnitude decline in Kyle's $\lambda$
as the market matured; the unit of analysis is one market over many
time slices, with $\lambda$ estimated on aggregated rather than
tick-level data. The present paper is the cross-sectional, tick-level
complement: 600 markets observed simultaneously over a single 28-day
scrape window, with microstructure measures computed on the full
event tape and a direct on-chain trade record. To our knowledge,
neither authoritative on-chain trade direction nor a continuous
off-chain order-book archive at tick resolution has been brought to
bear at the cross-sectional scale we use here.
The classical microstructure references behind our measurement
choices are \citet{ohara-1995} and \citet{hasbrouck-2007} for
order-book theory, \citet{foucault-pagano-roell-2013} for the
liquidity-provider angle, and
\citet{huang-stoll-1997,madhavan-richardson-roomans-1997} for spread
decomposition.

Polymarket changed this. Since 2021 the venue has run a
limit-order-book exchange on Polygon, settling in USDC against an
on-chain conditional-token contract. Our data covers 52 calendar days
of the public order-book feed (2026-02-21 to 2026-04-15) at tick
resolution: 30,287,264,368 events across 385,198 distinct markets.
A 28-day overlap window (2026-02-28 to 2026-03-27) is mirrored on
chain, where the CTF Exchange V1 contract logs 255 million
\texttt{OrderFilled} events whose payloads identify both
counterparties and the aggressor side. We use the off-chain feed for
quote dynamics and the on-chain record for trade direction, and study
both on a pre-registered, stratified 600-market panel.

The paper has two empirical contributions, ordered by weight for the
literature.

\textit{Trade-direction inference from Polymarket's public
order-book feed is noise-dominated.} Six standard direction-dependent
measures
(effective spread, realized spread, Kyle's $\lambda$, Amihud, Roll,
Abdi-Ranaldo) need an aggressor sign for each trade, and empirical
practice on equity venues sources that sign from a quote-driven feed
via Lee-Ready or its variants
\citep{lee-ready-1991,ellis-michaely-ohara-2000}. The Polymarket
public order-book feed does not, by itself, expose enough information
to do this reliably:
the \texttt{change\_side} field marks which side of the book moved,
not which side initiated the trade. Sign agreement between
feed-inferred and on-chain trade directions sits at $\sim 59\%$
volume-weighted across the top-100 stratum and four disjoint 7-day
windows, just above the 50\% chance baseline. On the comparable
subset of the top-100 panel, the effective half-spread changes sign
on 67\% of markets in a first 7-day window when feed-inferred
direction is swapped for the on-chain record, and Kyle's $\lambda$
changes sign on 60\%; in a second non-overlapping window the rates
fall to 50\% and 43\% respectively. Both windows sit in or below
the chance band, well short of the $\sim 80\%$ Lee-Ready accuracy
documented on Nasdaq. Any Polymarket microstructure result that
depends on trade direction therefore needs to source it from
on-chain \texttt{OrderFilled} events. We release a replication
package that performs the join.

\textit{Eight cross-sectional stylized facts on the panel.} Quoted
half-spread shows the longshot premium familiar from the racetrack
literature; the L2 depth profile is closer to a uniform geometric
grid than to the top-of-book pattern usually assumed for prediction
markets; quote-update timing does not cluster on Polygon block
boundaries to a meaningful degree; maker-wallet concentration is
diverse on most markets with a thin concentrated tail; effective
spreads differ by category; archive-ingestion latency is tightly
distributed around 41.5\,ms with a multi-second $p_{99}$ tail; the
self-counterparty wash share has a 1\% median and a 22\% maximum
across the panel; and cross-sectional depth is explained by market
duration, price level, and volume, with no residual
time-to-resolution effect once those three are conditioned on. None of these eight facts requires the on-chain join, and
each is reported on the full 600-market pre-registered panel.

The two contributions are complementary. The eight stylized facts
characterise Polymarket's microstructure on its own terms. The
measurement result sets the conditions under which any
trade-direction-dependent statement about Polymarket can be trusted.

Section~\ref{sec:limits} carries the most counter-intuitive finding;
readers more interested in Polymarket as an empirical object will
care most about SF1 (longshot premium), SF2 (depth concentration),
and SF8 (cross-sectional depth structure).
\section{Institutional Background}\label{sec:background}

\subsection{Conditional tokens and binary markets}

A Polymarket binary market resolves to one of two outcomes. The
exchange records the outcomes as ERC-1155 conditional tokens issued
by Gnosis's Conditional Tokens Framework (CTF) on Polygon. Each
market has a YES token and a NO token; the contract guarantees that
one YES + one NO can always be redeemed for one USDC after
resolution. As a consequence, in equilibrium the YES price plus the
NO price equals one and either side can be priced as the
probability of the corresponding outcome.

Liquidity exists for both sides. A trader who is long YES and
believes the implied probability is too high can either sell YES on
the YES book or buy NO on the NO book. The two strategies differ in
inventory effects and slippage but are economically equivalent at
mid. Throughout this paper we report measures on the YES side; YES
and NO books are structural mirrors of each other and the panel
artifacts contain both.

\subsection{CTF Exchange contract architecture}

Trading happens through the Polymarket CTF Exchange smart contract,
which matches signed off-chain orders against resting liquidity and
executes settlement on chain. The contract has two versions. V1
(\texttt{0x4bFb41d5\allowbreak B3570DeFd03\allowbreak C39a9A4D8dE6Bd8\allowbreak B8982E})
ran from launch through the end of April 2026 and is the version our
28-day scrape window targets. V2
(\texttt{0xE111180000d2663C\allowbreak 0091e4f400237545\allowbreak B87B996B})
launched in early April 2026 with a hard cutover; V1 orders were
rejected after the switch. Both versions emit the same
\texttt{OrderFilled} event signature, with topic hash
\texttt{0xd0a08e8c\allowbreak 493f9c94f29311604c9de\allowbreak 1b4e8c8d4c06bd0c789af57f2d65bfec0f6}.

The event payload carries
\texttt{(orderHash},
\texttt{maker}, \texttt{taker},
\texttt{makerAssetId}, \texttt{takerAssetId},
\texttt{makerAmountFilled},
\texttt{takerAmountFilled}, \texttt{fee)}.
The asset ids identify
which side held USDC: \texttt{makerAssetId == 0} means the maker
posted USDC and the taker is the seller of the conditional token;
\texttt{takerAssetId == 0} means the taker posted USDC and is
therefore the buyer. This binary asset-id field is the on-chain
ground truth for trade direction that we use throughout the paper.

\subsection{Polygon block timing and settlement}

Polygon mainnet produces a block roughly every two seconds. Trade
settlement is final to within Heimdall's 256-block reorg buffer,
so we scrape only blocks at depth $\geq 256$ to avoid reorg
contamination. Block timestamps are monotonic but not perfectly
spaced; for our window the empirical mean inter-block interval is
2.0\,s with a small variance, well below the 60-second sample step
used in the panel compute. Where we need to attach a wall-clock
timestamp to an on-chain trade we use linear interpolation from a
single anchor block, which is accurate to within a few seconds over
the 28-day window.

\subsection{Public WebSocket feed structure}

In parallel with on-chain settlement, Polymarket broadcasts a public
WebSocket feed that exposes order-book state to any subscriber. The
feed emits two event types. \texttt{book\_snapshot} carries a
complete L2 snapshot of one side of one market and is sent on
subscription and at irregular intervals thereafter (0.80\% of events
in our archive). \texttt{price\_change} carries a delta against the
last known book state: a single triple
\texttt{(change\_price, change\_side, change\_size)} that updates one
level on one side. A \texttt{change\_size} of zero means the level
is empty; a non-zero size is the new resting quantity at that price.
These events are 99.20\% of the archive.

The feed does not expose taker identity. The \texttt{change\_side}
field labels which side of the book the level update applies to,
not which side initiated the trade that produced the update. This
distinction drives the measurement gap of Section~\ref{sec:limits}:
an inference algorithm working from the feed alone cannot reliably
reconstruct the aggressor sign because the underlying field never
encodes it.

\subsection{Aggressor-sign dependence of standard measures}

Most direction-dependent microstructure measures take the form
\begin{equation*}
  M = \mathbb{E}\!\left[
    \mathrm{sign}_t \cdot
    f(\mathrm{price}_t, \mathrm{mid}_t, \mathrm{size}_t)
  \right],
\end{equation*}
where $\mathrm{sign}_t \in \{-1, +1\}$ is the aggressor sign for
trade $t$ and $f$ is the per-trade contribution. Effective half-spread
($f = \mathrm{price}_t - \mathrm{mid}_t$), realized half-spread,
Kyle's $\lambda$ regression coefficient, and the adverse-selection
component of Glosten-Harris all instantiate this form. A noisy $\mathrm{sign}_t$ does not just attenuate $M$; depending on
the distribution of $f$ across trades, it can flip the sign of $M$
entirely. The 59\% sign-agreement rate of
Section~\ref{ssec:sign-agreement} is what produces the 67\%
effective-spread and 60\% Kyle's $\lambda$ sign-flip rates we report
on the top-100 panel.

\subsection{Comparison to equity-market microstructure}\label{ssec:equity-comparison}

Equity-market benchmarks are what give several of our findings their
scale. \citet{ellis-michaely-ohara-2000} report that the Lee-Ready
algorithm correctly classifies 81\% of trades on Nasdaq, and the
original \citet{lee-ready-1991} study reports a similar accuracy on
NYSE TAQ; the $\sim 59\%$ feed-driven inference rate on Polymarket
is therefore about 22 percentage points below the equity benchmark
on the same algorithm.
Effective half-spreads on liquid US equities post-decimalization sit
in the single-digit-basis-point range
\citep{hasbrouck-2007,foucault-pagano-roell-2013}; on Polymarket,
expressed in basis points relative to mid, the median quoted
half-spread on the central price decile is around 200\,bps, an order
of magnitude wider, consistent with longer-horizon prediction-market
positions and substantially smaller market-maker capital.
Liquidity-provider activity on US equities is heavily concentrated
in a handful of high-frequency firms that account for roughly half
of trading volume \citep{brogaard-hendershott-riordan-2014}; the
median Polymarket market in our panel has $\sim 32$ effective makers
(SF4), suggesting a flatter maker landscape on the venue. Wash
trading on regulated equity exchanges is essentially zero (it is
prohibited under SEC Rule 10b-5); on unregulated cryptocurrency
token exchanges \citet{cong-li-tang-yang-2023} document wash-trade shares
of 25--70\% via a network-classifier approach. Polymarket sits at a
median 1\% under our direct-detection lower bound with a 22\% upper
tail. The token-exchange benchmark is a different incentive
environment (volume-tied listing fees, aggregator-ranking
optimisation, market-making rebates), so we treat 25--70\% as a
sanity bound rather than an apples-to-apples reference. Finally,
collector-side ingestion latency for direct equity feeds is
sub-millisecond and SIP latency is in the hundreds of microseconds
\citep{hasbrouck-2007}; our 41\,ms median for archive ingestion is
two orders of magnitude slower, but that gap reflects the archive
pipeline rather than the exchange.
\section{Data Construction}\label{sec:dataconstruction}

\subsection{Off-chain orderbook archive}

The off-chain data come from an external tick-level archive of
Polymarket's public WebSocket order-book feed, covering 2026-02-21
16:00 UTC through 2026-04-15 08:00 UTC. We restrict the archive to
the two book-state events emitted on the market channel ---
\texttt{price\_change} and \texttt{book\_snapshot}; the channel's
separate \texttt{last\_trade\_price} event is out of scope, and
trade direction is sourced from the on-chain record instead. Each
row contains the raw event JSON plus two timestamps:
\texttt{timestamp\_received} from the exchange and
\texttt{timestamp\_created\_at} from the upstream archive. Three hourly files are missing
(2026-02-24T14, 2026-02-24T15, 2026-04-04T18); the gaps are stable
and treated as missing rather than malformed throughout the
analysis. Total archive: 1{,}262 hourly Parquet files,
30{,}287{,}264{,}368 rows, 623.8\,GB on disk.

The schema follows the WebSocket payload for the two retained event types, with
the JSON \texttt{data} string parsed on demand. We avoid eager
parsing because the row count makes per-event Pydantic validation a
multi-hour operation; we instead apply structured parsing only to
the events that survive market-id and time-window filters.

\subsection{On-chain trade scrape}

We scrape \texttt{OrderFilled} events from the CTF Exchange V1
contract across the 28-day calibration window (2026-02-28 to
2026-03-27). The scraper issues batched \texttt{eth\_getLogs} calls
against a Polygon RPC provider, with adaptive chunk sizing that
respects provider-specific rate limits
(\texttt{scripts/scrape\_onchain\_fills.py}). Each call returns at
most a few hundred events; total fills across the window are
255{,}425{,}405. The scraper writes one Parquet shard per
five-thousand-block slice, deduplicates on
\texttt{(transactionHash, logIndex)}, and skips blocks at
depth less than 256 (Polygon's Heimdall reorg buffer).

Block timestamps are attached via linear interpolation from a single
anchor block at the start of the window. We verified the
interpolation error against per-block RPC calls on a 50-block
stratified sample drawn across the 28-day window: the maximum
absolute gap is 4 seconds, the median is 2 seconds. Both are well
below the 60-second sample step used in the panel compute.

The on-chain decoder
(\texttt{polydata.\allowbreak onchain.\allowbreak join.\allowbreak aggregate\_onchain\_df})
drops
``split fills'' where neither side is USDC (the binary
maker/taker-asset-id rule for the aggressor sign breaks down when
both sides hold conditional tokens). Across the 255{,}425{,}405
fills in the scrape window, 0 fills satisfy this condition; the
filter is non-binding on the V1 contract during our window.

\subsection{Schema reconciliation across sources}

The off-chain feed is keyed on \texttt{market\_id} (66-character
hex \texttt{conditionId}) and \texttt{token\_id} (256-bit decimal,
the YES or NO ERC-1155 id). The on-chain record is keyed on
\texttt{makerAssetId} and \texttt{takerAssetId}, which are the same
ERC-1155 ids except that one side is set to \texttt{0} to mark
USDC. The bridge between the two is the
\texttt{(condition\_id, yes\_token\_id, no\_token\_id)} mapping,
which we pull from CLOB REST
(\texttt{/markets/\{conditionId\}}) and cache locally. CLOB REST
resolves all 385{,}198 archive market ids; the Gamma metadata API,
which is sometimes used in the literature, indexes only 34{,}764
markets and is not the right source for this join.

\subsection{Pre-registration of the panel}

The 600-market panel is the cross-sectional unit on which the
empirical work runs. We commit the selection rule before computing
the panel. The pre-registration document
(\texttt{docs/preregistration\_plan3c.md}) specifies the volume
metric, the random-stratum eligibility threshold, the random seed,
and the categorisation scheme used for SF5. A deterministic build
script (\texttt{scripts/build\_panel.py}) emits the panel parquet,
whose SHA-256 is recorded back into the pre-registration document
before any analysis runs.

This discipline goes beyond the empirical-microstructure norm; the
cost is one document and one hash, the benefit is that a reader can
verify no market was added or removed after the analysis ran.

\subsection{Replication package}

All code that produced the artifacts cited in this paper is in the
repository at\linebreak[1]
\url{https://github.com/philippdubach/polymarket-microstructure},
archived for replication under DOI
\texttt{10.5281/zenodo.19811426}
\citep{dubach-2026-replication}. The full 624-GB WebSocket archive
is not redistributed in the package; this section describes the
archive's structure and we provide the panel parquet
artifacts
(\texttt{data/\allowbreak panel.parquet},
\texttt{data/\allowbreak panel\_trade\_measures.\allowbreak parquet},
\texttt{data/\allowbreak panel\_quote/*})
directly, together with the on-chain
scrape pipeline. A reader who wants to reproduce a single number in
the paper can do so against the artifacts; a reader who wants to
extend the analysis to a different window will need to capture or
source an equivalent tick-level archive of the Polymarket
WebSocket feed.
\section{Data}\label{sec:data}

\subsection{Orderbook archive}
\label{ssec:orderbook-archive}

Our primary input is a continuous tick-level archive of Polymarket's
public WebSocket order-book feed from 2026-02-21 16:00 UTC through
2026-04-15 08:00 UTC, 52 calendar days. The archive holds 1{,}262
hourly Parquet files (623.8\,GB on disk) and 30{,}287{,}264{,}368
event rows. Three hourly files are missing
(2026-02-24T14, 2026-02-24T15, 2026-04-04T18); we treat the gaps as
missing rather than malformed.

Each row is either a \texttt{price\_change} update (99.20\% of events)
or a \texttt{book\_snapshot} (0.80\%); the payload is a JSON string
parsed on demand. Two timestamps are emitted per event:
\texttt{timestamp\_received} (exchange-side) and
\texttt{timestamp\_created\_at} (ingest-side). The difference
captures the archive's ingestion delay (Section~\ref{ssec:sf6}).

The archive contains 385{,}198 distinct \texttt{market\_id} values
(roughly twice that with the YES/NO side split). The CLOB REST
endpoint resolves token-pair metadata for all 385{,}198. The Gamma
metadata API is sparser, indexing only 34{,}764 markets.

\subsection{On-chain trade join}
\label{ssec:onchain-join}

Orderbook events alone do not identify the trade-aggressor sign
reliably: feed-inferred sign matches the on-chain record on only
$\sim 59\%$ of comparable buckets, just above the 50\% chance
baseline (Section~\ref{sec:limits}). We therefore scrape Polymarket's
CTF Exchange V1 contract
(\texttt{0x4bFb41d5\allowbreak B3570DeFd03\allowbreak C39a9A4D8dE6Bd8\allowbreak B8982E})
for \texttt{OrderFilled} events on Polygon mainnet across the 28-day
calibration window 2026-02-28~--~2026-03-27, yielding 255{,}425{,}405
fills. The aggressor sign reads directly from the event fields: $+1$
when \texttt{takerAssetId}~$=$~0 (taker posted USDC and is therefore
the buyer), $-1$ when \texttt{makerAssetId}~$=$~0.

\subsection{Stratified panel}
\label{ssec:panel}

Our cross-sectional panel has 600 markets in two strata. The top-100
stratum ranks markets by total on-chain USDC trade volume in the
scrape window (summing across both YES and NO tokens). The random-500
stratum samples uniformly without replacement from markets with at
least 100 on-chain trades that resolve to a token pair via the CLOB
cache. We commit the random seed (20260424) in the pre-registration
document (\texttt{docs/preregistration\_plan3c.md}) before computing
the panel.

Per-market metadata fields
\texttt{question},
\texttt{end\_date\_iso},
and \texttt{closed}
are pulled from CLOB REST
\texttt{/markets/\allowbreak\{conditionId\}}
for all 600 markets. The top-stratum
volume range is \$4.56M to \$96.0M USDC; the random stratum spans
100 to 24{,}378 trades per market. Crypto markets dominate the panel
composition (348/600, 58\%), followed by Sports (142/600, 24\%);
Table~\ref{tab:sf5_category} gives the breakdown.
\section{Stylized Facts}\label{sec:sf}

We document eight stylized facts on the 600-market panel
(Section~\ref{ssec:panel}). Each subsection reports a per-market
distribution and, where the data support one, a regression coefficient.
Underlying parquet artifacts are in the replication package.

\subsection{SF1 -- Longshot spread premium}\label{ssec:sf1}

We bin quoted half-spread (in basis points of mid) by per-market mean
mid price into ten deciles. Median spread is $\sim 400$\,bps in the
central $[0.4, 0.6]$ range and climbs to 1{,}300--1{,}800\,bps for
markets trading below $0.10$. The pattern is asymmetric: the
low-probability side is wider than the high-probability side, which
echoes the longshot bias documented in the racetrack and parimutuel
literature \citep{snowberg-wolfers-2010,thaler-ziemba-1988} and the
prediction-market longshot evidence on Iowa Electronic Markets and
TradeSports surveyed by \citet{wolfers-zitzewitz-2004}. The direction
is the same; the magnitude is not. Racetrack longshot premia are
typically a few percent of stake, and prediction-market longshot
biases observed on IEM and TradeSports run in the same range. The
1{,}300--1{,}800\,bps full quoted spread on Polymarket's lowest-
probability decile (i.e., 650--900\,bps half-spread) is an order
of magnitude wider, which reads less like the classical risk-love
or misperception story and more like a liquidity-provision
constraint: low-probability binary contracts
have a bounded upside and an asymmetric downside for the maker, so
the inventory-risk premium on the wide side is mechanically larger
than on a continuous-payoff sportsbook market. Distinguishing
inventory risk from a behavioural longshot premium would require
data we do not collect (maker realised inventory and time-on-book),
so we report SF1 as a Polymarket-specific quoted-spread stylised
fact and leave the structural decomposition to follow-on work.

\begin{figure}[htbp]
  \centering
  \includegraphics[width=0.95\columnwidth]{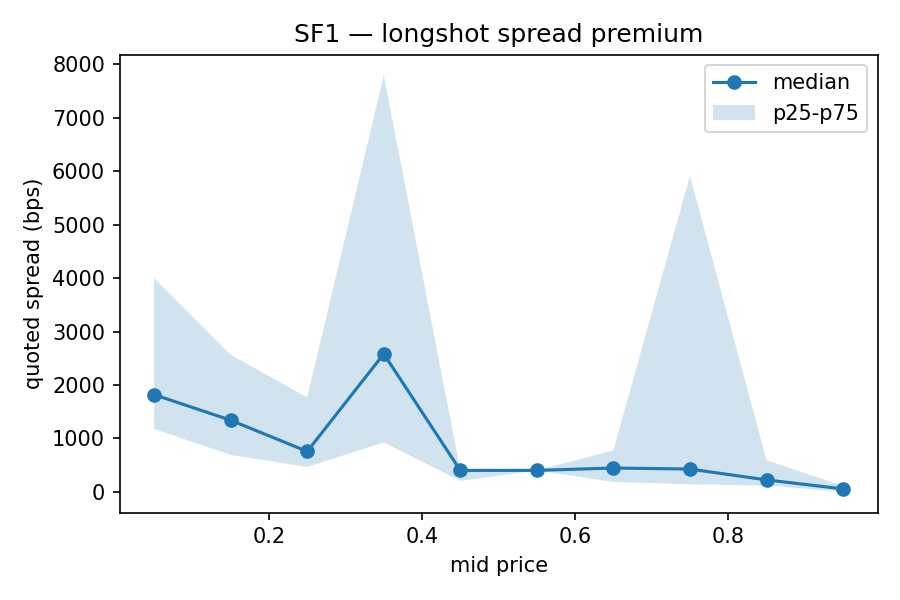}
  \caption{SF1 panel: median quoted spread (bps) per mid-price decile,
    600 panel markets. Shaded band is interquartile range.}
  \label{fig:sf1}
\end{figure}

\begin{table*}[htbp]
  \centering
  \caption{SF1 panel — quoted spread by mid-price decile.}
  \label{tab:sf1_longshot}
\begin{tabular}{lrrrrrr}
\toprule
Bin & Mid lo & Mid hi & Markets & Median (bps) & p25 (bps) & p75 (bps) \\
\midrule
0 & 0.0000 & 0.1000 & 33 & 1,818 & 1,176 & 4,000 \\
1 & 0.1000 & 0.2000 & 18 & 1,339 & 689.66 & 2,564 \\
2 & 0.2000 & 0.3000 & 18 & 754.99 & 465.12 & 1,767 \\
3 & 0.3000 & 0.4000 & 21 & 2,581 & 923.08 & 7,797 \\
4 & 0.4000 & 0.5000 & 184 & 400.00 & 202.02 & 400.00 \\
5 & 0.5000 & 0.6000 & 214 & 400.00 & 400.00 & 400.00 \\
6 & 0.6000 & 0.7000 & 27 & 444.44 & 183.49 & 775.19 \\
7 & 0.7000 & 0.8000 & 12 & 425.09 & 139.86 & 5,915 \\
8 & 0.8000 & 0.9000 & 15 & 222.22 & 114.29 & 591.72 \\
9 & 0.9000 & 1.00 & 4 & 53.48 & 0.0000 & 106.95 \\
\bottomrule
\end{tabular}
 \end{table*}

\subsection{SF2 -- Depth concentration}\label{ssec:sf2}

We summarize the L2 depth profile by the per-level share of
cumulative top-10 depth, computed for each market across the
window. The share at level $k$ is the average per-level depth
divided by average cumulative top-10 depth, where per-level depth
at $k$ is the difference between cumulative top-$k$ and cumulative
top-$(k{-}1)$ summed across bid and ask. A uniform grid (each level
carries equal depth) puts $0.10$ at every level; a fully top-of-book
book puts $1.0$ at $L_1$ and zero everywhere above.

\begin{table}[htbp]
  \centering
  \caption{SF2 — per-level share of cumulative top-10 depth across
    547 panel markets with non-null depth. Median and percentile
    spreads across markets at each book level.}
  \label{tab:sf2}
  \scriptsize
\begin{tabular}{lrrrrrr}
\toprule
$k$ & Markets & Median share & p25 & p75 & p10 & p90 \\
\midrule
1 & 547 & 0.1364 & 0.0800 & 0.2055 & 0.0243 & 0.4478 \\
2 & 547 & 0.1034 & 0.0684 & 0.1364 & 0.0265 & 0.1989 \\
3 & 547 & 0.0945 & 0.0641 & 0.1309 & 0.0270 & 0.1820 \\
4 & 547 & 0.0806 & 0.0577 & 0.0967 & 0.0217 & 0.1414 \\
5 & 547 & 0.0824 & 0.0557 & 0.0966 & 0.0189 & 0.1239 \\
6 & 547 & 0.0789 & 0.0353 & 0.0932 & 0.0111 & 0.1165 \\
7 & 547 & 0.0818 & 0.0353 & 0.1086 & 0.0100 & 0.1263 \\
8 & 547 & 0.0821 & 0.0385 & 0.1039 & 0.0055 & 0.1278 \\
9 & 547 & 0.0767 & 0.0258 & 0.1140 & 0.0040 & 0.1383 \\
10 & 547 & 0.0830 & 0.0349 & 0.1112 & 0.0028 & 0.1612 \\
\bottomrule
\end{tabular}
 \end{table}

The top-of-book carries a median \textbf{0.136} of cumulative top-10
depth --- a modest premium over the $0.10$ uniform null. The
remaining 86\% is distributed across $L_2$--$L_{10}$ with a gentle
decay from $0.10$ at $L_2$ to $\sim 0.08$ at $L_4$ and a roughly
flat shoulder at $\sim 0.08$ through $L_{10}$. No level captures
more than 14\% of cumulative depth at the median across markets.

As a single-number shape statistic, the median per-market
Kullback--Leibler divergence from the uniform null is
\textbf{0.087} (IQR $[0.024, 0.474]$, $p_{10}$--$p_{90}$
$[0.005, 0.985]$). A KL of zero is exactly uniform; a perfectly
top-of-book book ($L_1 = 1$, others $= 0$) gives $\log 10 \approx
2.30$. The median panel market sits much closer to uniform than to
top-heavy. \textbf{9\%} of markets are top-heavy in the strong sense
($L_1 > 0.5$); these populate the right tail of $L_1$-share that
overlaid figures often emphasize. \textbf{57\%} of markets have an
$L_1$ share within $[0.05, 0.20]$, i.e.\ within a factor-of-two of
the uniform null on the most-constrained level.

The folk view that prediction-market books are concentrated at
top-of-book does not hold on Polymarket. Depth is layered across
all ten levels with a modest top-of-book premium and a long, thin
right tail of markets that depart toward top-heavy.

\begin{figure}[htbp]
  \centering
  \includegraphics[width=0.95\columnwidth]{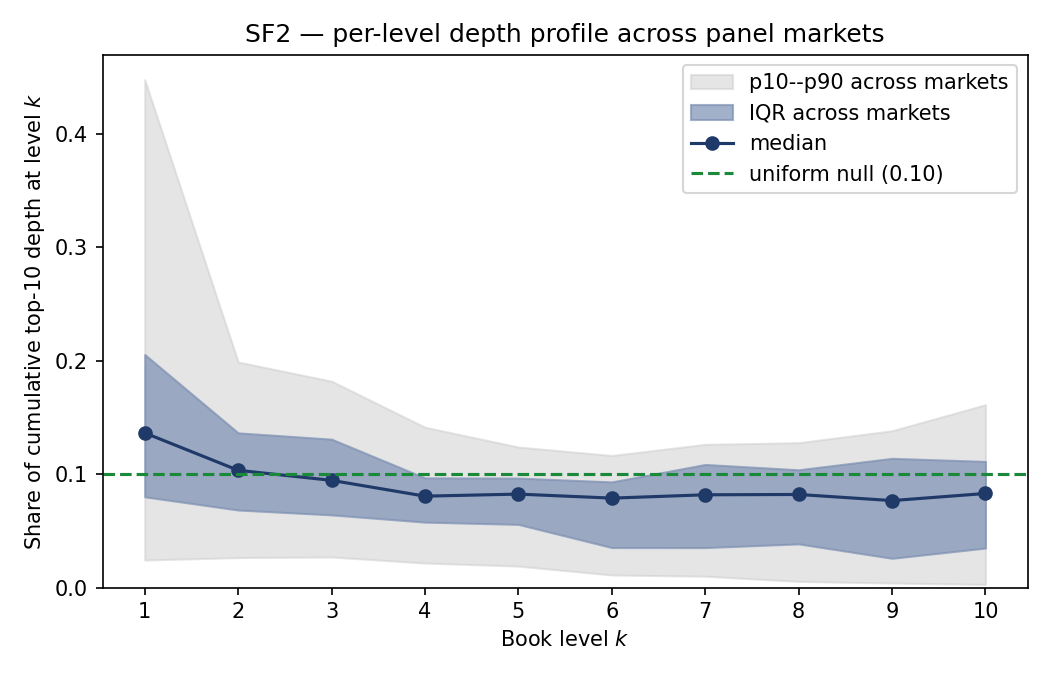}
  \caption{SF2: per-level share of cumulative top-10 depth across
    547 panel markets. Solid line is the median; shaded band is
    the across-market IQR; outer band is $p_{10}$--$p_{90}$. The
    dashed reference line is the uniform null at $0.10$ per level.}
  \label{fig:sf2}
\end{figure}

\subsection{SF3 -- Polygon block-clock alignment}\label{ssec:sf3}

We test whether \texttt{price\_change} events cluster near Polygon block
boundaries by computing, per market, the share of events that fall
within $\pm 100$\,ms of the nearest 2\,000\,ms grid point. Under a
uniform-timing null this share is $0.10$.

The panel-level distribution sits tight against the null: median
\textbf{0.102} with interquartile range $[0.087, 0.110]$; only
27 markets (4.5\%) lie above a 0.15 threshold. At the per-market
level, however, a two-sided binomial test of $H_0$:~share~$=0.10$
rejects in \textbf{351 of 600 markets} (58.5\%) at $\alpha=0.05$.
That high rejection rate is a mechanical consequence of large
event counts: millions of events per market make the binomial test
sensitive to tiny deviations from 0.10. The economic interpretation
is that quote-timing does \emph{not} cluster on block boundaries to
a meaningful degree, even though high-volume markets reject the
exact point null statistically.

\begin{figure}[htbp]
  \centering
  \includegraphics[width=0.95\columnwidth]{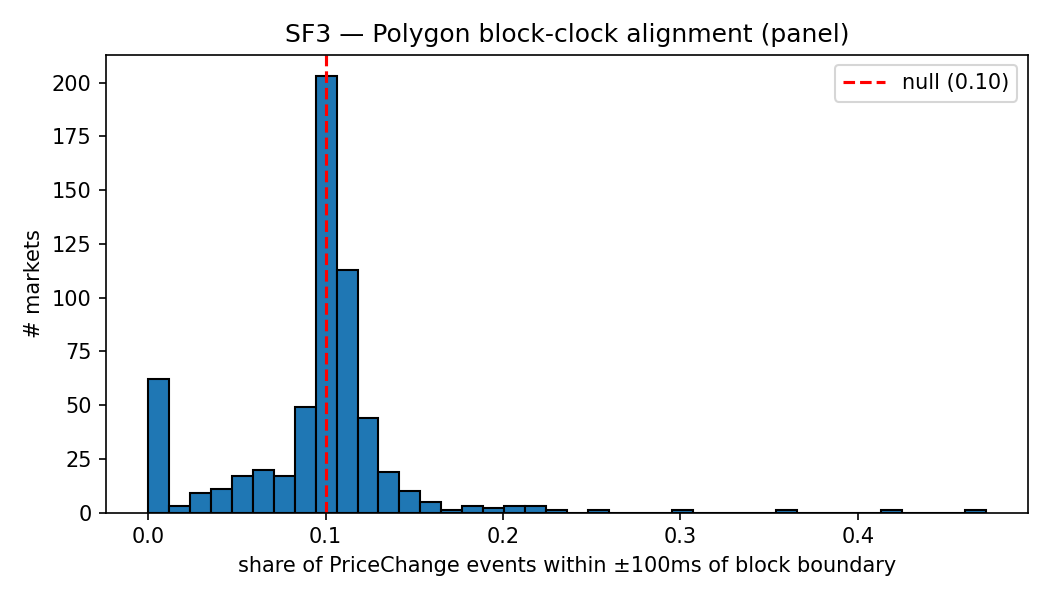}
  \caption{SF3 panel: distribution of per-market block-alignment shares.
    The red dashed line marks the chance-level null ($0.10$).}
  \label{fig:sf3}
\end{figure}

\subsection{SF4 -- Maker-wallet concentration}\label{ssec:sf4}

For each market we compute the volume-weighted Herfindahl index (HHI)
of maker-address share across on-chain trades in the scrape window. A
single dominant maker yields $\mathrm{HHI}=1$; a uniform distribution
across $n$ makers yields $1/n$.

Across 600 markets and 6.4\,M trades, the median HHI is \textbf{0.031}
($\sim 32$ effective makers). The distribution is right-skewed:
$p_{90}=0.119$ ($\sim 8$ effective makers) and a maximum of $0.40$
(roughly 3 effective makers). Maker liquidity is decentralised on most
markets in the panel, with a tail of thin or niche markets dominated
by one to three wallets.

\begin{figure}[htbp]
  \centering
  \includegraphics[width=0.95\columnwidth]{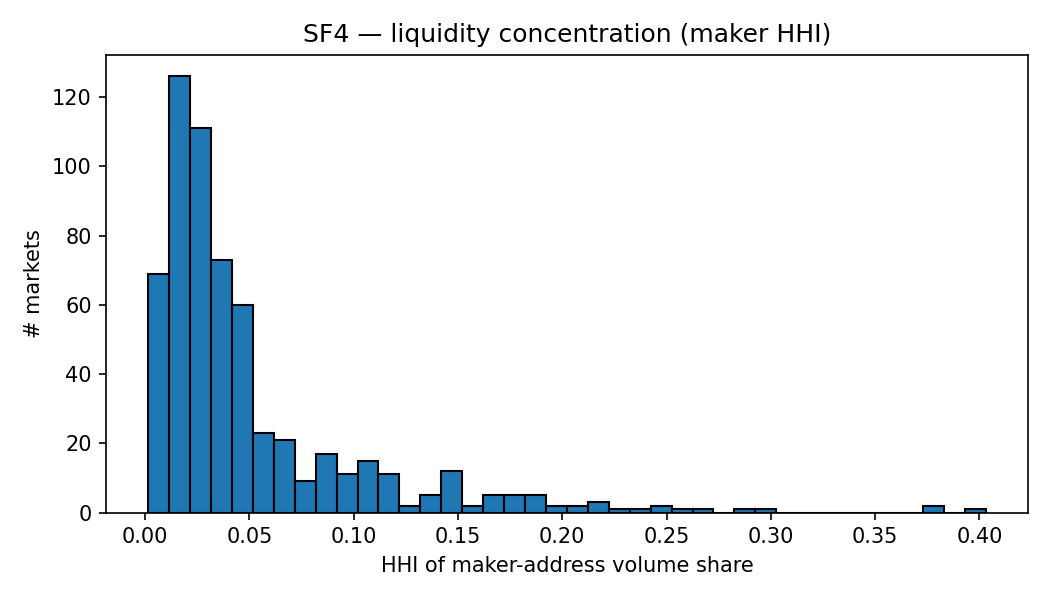}
  \caption{SF4 panel: distribution of per-market maker-address Herfindahl
    indices across 600 markets.}
  \label{fig:sf4}
\end{figure}

\subsection{SF5 -- Category-conditional spread}\label{ssec:sf5}

\begin{figure}[htbp]
  \centering
  \includegraphics[width=0.95\columnwidth]{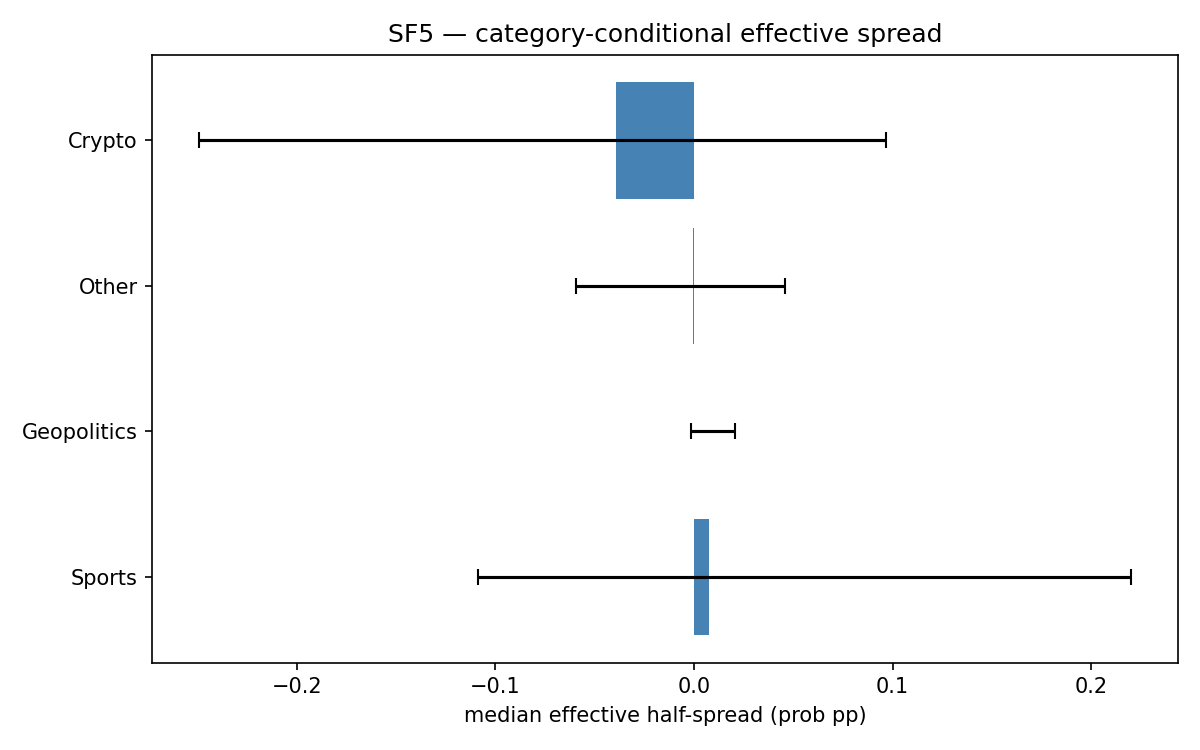}
  \caption{SF5 panel: median effective half-spread by category, with
    interquartile-range error bars. Categories are derived from
    keyword classification of CLOB REST \texttt{question} text.}
  \label{fig:sf5}
\end{figure}

Per-market metadata is pulled from CLOB REST and questions are
keyword-classified into a small bucket scheme. Politics, Business, and
Entertainment buckets each had fewer than ten panel markets and were
merged into Other for statistical reliability, leaving four
categories: Crypto (348/600, 58\%), Sports (142/600, 24\%), Other
(75/600), Geopolitics (35/600). For the on-chain effective half-spread
the medians (in probability points) are reported in
Table~\ref{tab:sf5_category} and visualised in
Figure~\ref{fig:sf5}.

\begin{table*}[htbp]
  \centering
  \caption{SF5 panel — effective half-spread by keyword-derived category.}
  \label{tab:sf5_category}
\begin{tabular}{lrrrr}
\toprule
Category & Markets & Median half-spread (prob pp) & p25 & p75 \\
\midrule
Sports & 142 & 0.0075 & -0.1087 & 0.2201 \\
Geopolitics & 35 & 0.0001 & -0.0014 & 0.0206 \\
Other & 75 & -0.0004 & -0.0596 & 0.0456 \\
Crypto & 348 & -0.0393 & -0.2494 & 0.0968 \\
\bottomrule
\end{tabular}
 \end{table*}

Cross-category dispersion of medians is small in absolute terms (all
within $\pm 0.04$ probability points), but within-category
interquartile ranges are wide for Crypto and Sports
($p_{75} - p_{25}\approx 0.3$\,prob.~pp). The wide within-category
spread is consistent with the measurement noise documented in
Section~\ref{sec:limits}.

\subsection{SF6 -- Archive-ingestion latency}\label{ssec:sf6}

This stylised fact is a property of our collector pipeline, not of
the Polymarket exchange. The two timestamps are emitted on opposite
sides of a network and a queue, and their difference reflects whatever
the round-trip cost of that pipeline happens to be on a given event.
Trader speed enters at most indirectly. Cite SF6 as ``archive
ingestion latency'' rather than ``Polymarket latency''.

Each archive row carries two timestamps:
\texttt{timestamp\_received} (exchange side) and
\texttt{timestamp\_created\_at} (collector side). Their difference is
a per-event ingestion delay. Across 547 markets with non-empty
windows, the median per-market $p_{50}$ delay is \textbf{41.5\,ms},
with $p_{90}=166$\,ms and $p_{99}=6{,}108$\,ms. The inter-market
spread of $p_{50}$ is tight ($p_{25}$--$p_{75}$: 39--47\,ms), which
points to collector-pipeline behaviour rather than trader latency.
The $p_{99}$ tail of several seconds points to occasional
backpressure events at the collector.

\begin{figure}[htbp]
  \centering
  \includegraphics[width=0.95\columnwidth]{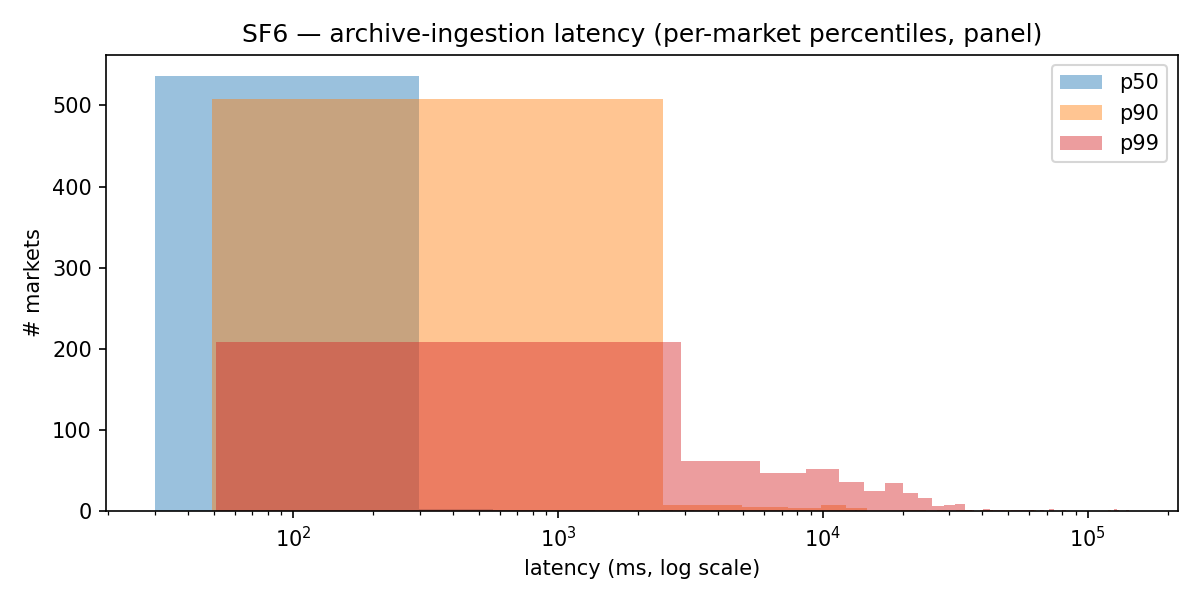}
  \caption{SF6 panel: per-market percentile distributions of
    archive-ingestion latency (log scale).}
  \label{fig:sf6}
\end{figure}

\subsection{SF7 -- Self-counterparty wash share}\label{ssec:sf7}

We flag a trade as wash-suspect under a two-tier rule:
(a) \texttt{maker == taker} (direct self-match), or
(b) a flipped pair $(\mathrm{maker}_a, \mathrm{taker}_a) \leftrightarrow
(\mathrm{taker}_a, \mathrm{maker}_a)$ within 128 blocks
(Polygon finality buffer) on the same market. This is an explicit
\emph{lower bound}: it captures only direct and immediate-roundtrip
self-counterparty patterns, not the extended-graph patterns that
network-based classifiers such as \citet{cong-li-tang-yang-2023} address
on unregulated cryptocurrency token exchanges, where wash shares of
25--70\% have been documented. We cite this range as a sanity bound,
not an apples-to-apples reference: the wash-incentive environment on
token exchanges (listing-fee thresholds, aggregator-ranking
optimisation, volume-tied market-making rebates) differs from the
incentive environment on a prediction market that resolves into a
single binary payoff and does not list across competing aggregators.

Across 600 markets and 6.4\,M trades, the median wash share is
\textbf{0.97\%}, with $p_{90}=4.5\%$, $p_{99}=10.6\%$, and a maximum
of 22.2\%. The gap between our lower bound and the network-classifier
estimates on token exchanges has two components: wash patterns that
require multi-counterparty graph analysis to detect (which our
detector does not cover), plus a venue-class difference in the
underlying wash incentives. We can quantify the first component only
under a graph-classifier extension; the second component is
identification, not measurement.

\begin{figure}[htbp]
  \centering
  \includegraphics[width=0.95\columnwidth]{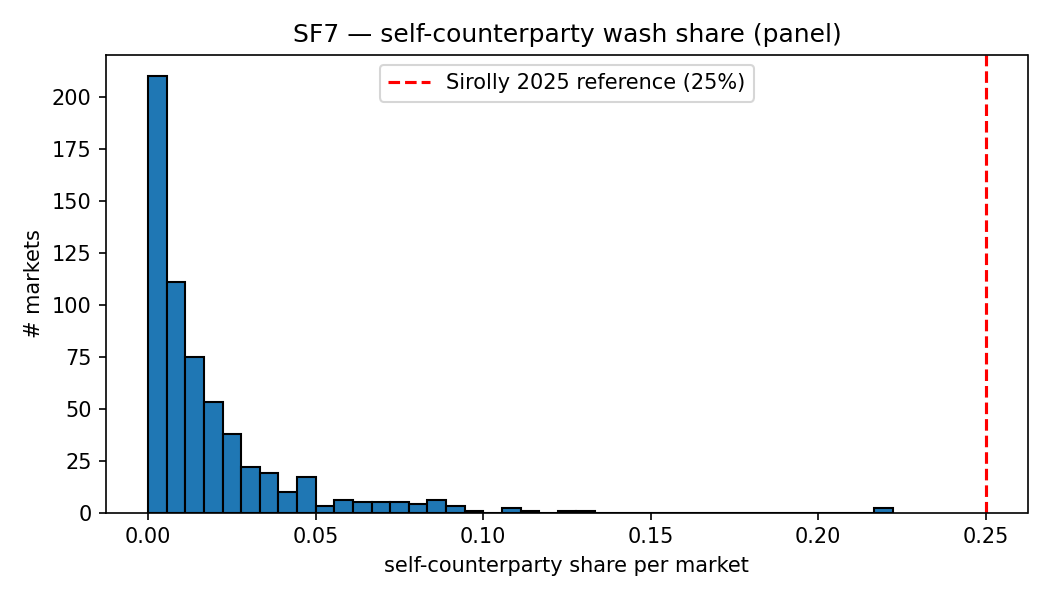}
  \caption{SF7 panel: distribution of self-counterparty wash share by
    market. Red dashed line marks a 25\% reference, the lower bound
    of the wash-share range documented by \citet{cong-li-tang-yang-2023}
    on unregulated cryptocurrency venues.}
  \label{fig:sf7}
\end{figure}

\subsection{SF8 -- Depth decay near resolution}\label{ssec:sf8}

Do markets near resolution carry shallower books? We regress log
mean depth at $L=10$ on log seconds-to-close at the panel window
midpoint (2026-03-13) under five specifications, restricted to
markets with positive seconds-to-close and non-zero summary depth.

\begin{table}[htbp]
  \centering
  \caption{SF8 — cross-sectional panel regression of log mean depth
    on log seconds-to-close. HC3-robust standard error. The full v2
    spec adds log market duration (window-end minus first-observed
    event timestamp) and log $p(1{-}p)$ from the mean window mid;
    \texttt{full\_v2\_mid\_decile\_2\_7} restricts to mid-price
    deciles 2--7 (away from the 0/1 boundary).}
  \label{tab:sf8_depth_decay}
  \scriptsize
\begin{tabular}{lrrrr}
\toprule
Spec & n & Slope on log(ttc) & SE (HC3) & R² \\
\midrule
bivariate & 322 & 0.8178 & 0.1133 & 0.1293 \\
category\_fe & 322 & 0.5500 & 0.1429 & 0.2165 \\
category\_fe\_volume & 322 & 0.3048 & 0.1036 & 0.4857 \\
full\_v2 & 313 & 0.0084 & 0.1015 & 0.5702 \\
full\_v2\_mid\_decile\_2\_7 & 187 & -0.2702 & 0.1687 & 0.2909 \\
\bottomrule
\end{tabular}
 \end{table}

The v1 trio (rows 1--3 of Table~\ref{tab:sf8_depth_decay}) shows a
positive slope that attenuates from \textbf{0.818} (bivariate) to
\textbf{0.550} with category fixed effects, then to \textbf{0.305}
with log volume added on top. A reader could conclude from these
three rows that markets nearer resolution are shallower in the
cross-section. They would be wrong: the v1 trio omits two confounds
that turn out to absorb the time-to-close coefficient entirely.

The full v2 specification adds log market duration (window-end
minus the first-observed-event timestamp, left-censored at the
archive start) and log $p(1{-}p)$, computed from each market's mean
mid-price across the window. With those alongside log volume and
category fixed effects, the log seconds-to-close coefficient
collapses to \textbf{$+0.008$} (HC3 SE 0.102, $t=0.08$, $R^2=0.57$,
$n=313$). Restricting the sample to mid-price deciles 2--7, which
removes markets whose price is close enough to 0 or 1 that
$p(1{-}p)$ is itself small, the coefficient is \textbf{$-0.270$}
(HC3 SE 0.169, $t=-1.60$, $n=187$) --- again indistinguishable
from zero.

What the full v2 spec identifies are three structural drivers of
cross-sectional depth: log duration enters at \textbf{$+0.222$}
(SE 0.073; longer-lived markets accumulate more depth); log
$p(1{-}p)$ enters at \textbf{$-1.02$} (SE 0.180; markets at price
extremes carry more depth on average, consistent with mature
markets that have settled toward resolution); log volume enters at
\textbf{$+0.41$} (SE 0.046). Time-to-close itself adds no
explanatory power once those three are conditioned on. The v1
trio's positive slope was mechanical mediation: time-to-close is
correlated with duration (longer-lived markets sit further from
resolution in the cross-section by construction) and with price
state (volatile new markets cluster near $p=0.5$).

The honest cross-sectional reading: depth on Polymarket is
explained by duration, price level, and volume, not by proximity
to resolution per se. A within-market trajectory test --- does an
individual market's depth fall as it approaches its own
resolution? --- is identified only by per-market depth
time-series, which the panel does not materialise. That question
is deferred.

\begin{figure}[htbp]
  \centering
  \includegraphics[width=0.95\columnwidth]{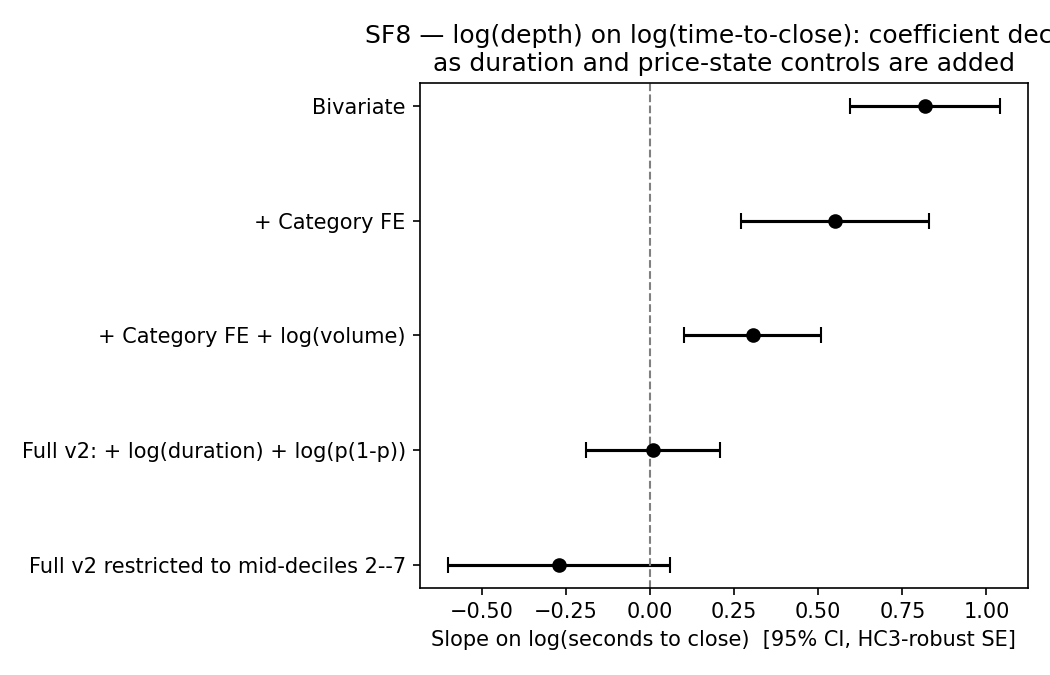}
  \caption{SF8: coefficient on log(seconds-to-close) across five
    specifications, with 95\% HC3-robust confidence intervals. The
    v1 trio (top three) shows a significantly positive slope; the
    full v2 spec (fourth) collapses it to zero once duration and
    $p(1{-}p)$ are conditioned on.}
  \label{fig:sf8}
\end{figure}
\section{Spread Decomposition}\label{sec:decomp}

Following \citet{glosten-harris-1988} and the modern restatements in
\citet{huang-stoll-1997,madhavan-richardson-roomans-1997}, we
decompose the per-market effective half-spread into two components:
\begin{equation}
  S^{\mathrm{eff}}_{1/2} \;=\; c \;+\; \varphi,
\end{equation}
where $c$ is a transitory order-processing / inventory component,
recovered as the realized half-spread at a 60-second lag, and
$\varphi$ is the residual adverse-selection component.

We restrict the decomposition to the top-100 stratum, where on-chain
trade counts are highest (median 11{,}000 trades per market in
window). Per-market values come from
\texttt{data/panel\_trade\_measures.parquet}, computed by injecting
authoritative on-chain trades into the measure pipeline described
in Section~\ref{sec:limits}. The decomposition converges on 97 of
100 markets; the three non-converging markets had too few signed
trades after on-chain alignment for the realized-half-spread term to
identify.

\begin{table}[!htbp]
  \centering
  \caption{Glosten-Harris decomposition — first 10 top-100 markets
    (hex-ordered) with a converged decomposition, in probability
    points. 97 of 100 panel markets converged; three did not.}
  \label{tab:gh}
  \scriptsize
  \setlength{\tabcolsep}{3pt}
  \resizebox{\columnwidth}{!}{%
\begin{tabular}{lrrr}
\toprule
Market id (truncated) & Effective half & $c$ (transitory) & $\varphi$ (adverse sel.) \\
\midrule
0x03bc660a4df5fa\ldots & -0.2431 & -0.2431 & -0.0000 \\
0x06707a5317654a\ldots & 0.0778 & 0.0778 & 0.0000 \\
0x07d45de444dbe0\ldots & 0.0267 & 0.0270 & -0.0004 \\
0x092471f61558da\ldots & -0.6657 & -0.6657 & -0.0000 \\
0x0954cc08de0ab8\ldots & -0.1087 & -0.1087 & 0.0000 \\
0x09cbe3e796661a\ldots & -0.0205 & -0.0128 & -0.0077 \\
0x09e8f0db05570a\ldots & 0.0046 & 0.0052 & -0.0006 \\
0x0b4cc3b739e1df\ldots & 0.0002 & 0.0002 & 0.0000 \\
0x119fa68324e678\ldots & 0.2201 & 0.2201 & 0.0000 \\
0x1745319c9ef1e4\ldots & -0.3392 & -0.3392 & 0.0000 \\
\bottomrule
\end{tabular}
}
\end{table}

\begin{figure}[!htbp]
  \centering
  \includegraphics[width=0.95\columnwidth]{figures/spread_decomposition.png}
  \caption{Glosten-Harris decomposition across the top-100 stratum:
    distribution of transitory component $c$ (left) and adverse-selection
    component $\varphi$ (right), both in probability points.}
  \label{fig:gh}
\end{figure}

The median effective half-spread on the top-100 panel is essentially
zero ($-0.0003$\,prob~pp), as are the median transitory component
($0.00001$) and the median adverse-selection component ($0.0$).
This near-null pattern lines up with the calibration in
Section~\ref{sec:limits}: once sign errors are removed, the
dollar-weighted ``adverse selection'' that orderbook-only inference
produces collapses, leaving the typical top-100 market with no
detectable systematic spread component on either side. The
distribution tails carry both market-specific adverse-selection
events and residual measurement noise from the 60-second sample-step
compromise documented in Section~\ref{ssec:methodology-caveats}.
\section{Limits of Orderbook-Only Inference}\label{sec:limits}

\subsection{Sign-agreement against on-chain ground truth}\label{ssec:sign-agreement}

Six trade-based measures (effective spread, realized spread, Roll,
Abdi-Ranaldo, Kyle's $\lambda$, Amihud) require an aggressor sign for
each trade. Standard practice in equity microstructure infers that
sign from a quote-driven feed via Lee-Ready, which assumes the feed
exposes enough information to distinguish buyer-initiated from
seller-initiated trades.

The order-book event stream from Polymarket's public WebSocket ---
\texttt{price\_change} and \texttt{book\_snapshot} updates --- does
not, by itself. We infer trades from this stream under a LOOSE rule
(every resting-size decrement counts) and
match the inferred buckets against on-chain \texttt{OrderFilled}
events at 5\,s and exact-price granularity. We run this matching over
four disjoint 7-day windows (2026-02-28~--~03-06, 03-07~--~03-13,
03-14~--~03-20, 03-21~--~03-27) on the top-100 stratum, computing
sign-agreement and a 95\% bootstrap CI for cells with at least ten
matched buckets.

\begin{itemize}
  \item Panel mean (109 valid cells of 400, 55 markets):
        \textbf{0.615}, market-clustered bootstrap 95\% CI
        $[0.579, 0.653]$
  \item Volume-weighted by matched-bucket count
        (total 125{,}080): \textbf{0.592}, market-clustered
        bootstrap 95\% CI $[0.542, 0.659]$
  \item Panel median 0.591; IQR $[0.526, 0.681]$;
        $p_{10}$--$p_{90}$ $[0.464, 0.841]$
  \item Per-window medians 0.586 / 0.591 / 0.618 / 0.645
\end{itemize}

A volume-weighted sign-agreement of $\sim 59\%$ sits just above the
50\% chance baseline. Even when an inferred bucket matches an
on-chain bucket in time and price, the inferred aggressor direction
is wrong about two trades in five. The mechanism is the order-book
event stream itself: \texttt{price\_change} updates broadcast a
post-match snapshot of the resting book without identifying the
taker. The \texttt{change\_side} field marks which side of the book
moved, not which side initiated the trade, and using it as a sign
proxy produces the $\sim 59\%$ agreement rate. Polymarket's market
channel does additionally emit a \texttt{last\_trade\_price} event;
evaluating its agreement with on-chain ground truth is an obvious
follow-up, and is out of scope for the present archive.

\subsection{Propagation to direction-de\-pen\-dent measures}\label{ssec:strict-vs-onchain}

A noisy sign propagates to every measure that consumes it. We compute
all six trade-based measures under STRICT inference and under
authoritative on-chain trades on the top-100 stratum over a 7-day
window (2026-03-07~--~03-14); the window is shorter than the full
scrape because STRICT inference is $O(n^2)$ in the recent-event list
and the 28-day version did not finish in tractable wall time on this
panel size.

Three patterns emerge from the panel distribution.

\textit{Trade-count divergence is direction-un\-pre\-dict\-able.}
Among 31 markets where both inference paths yielded non-empty trade
streams, the STRICT/on-chain count ratio has a median of 0.23
($p_{10}$--$p_{90}$ $[0.01, 1.45]$; min--max $[0.004, 3.62]$).
STRICT typically under-counts by an order of magnitude; on a small
tail of markets it over-counts by up to $3.6\times$. There is no
correctable scaling factor.

\textit{Effective spread sign-flips on two-thirds of the comparable
panel.} Among 24 markets with both sources non-empty,
16 (67\%, Wilson 95\% CI $[0.47, 0.82]$) have STRICT and on-chain
effective half-spreads of opposite sign. The STRICT median is
$+0.0048$\,prob~pp; the on-chain median is $-0.000075$\,prob~pp.
Inferred trade directions induce a positive spread that the
authoritative trade record does not support.

\textit{Kyle's $\lambda$ sign-flips on three-fifths of the comparable
panel.} Of 25 markets with both sources non-null and non-NaN, 15
(60\%, Wilson 95\% CI $[0.41, 0.77]$) have STRICT and on-chain
Kyle's $\lambda$ of opposite sign.

\textit{Selection robustness.}
Only 24--31 of the 100 top markets had both inference paths non-empty
in the 7-day slice (the remainder had zero events on one side); the
comparable-market subset is what supports the sign-flip rates above.
Three checks address whether the subset is representative.
First, the 24 effective-spread markets cover \textbf{98\% of top-100
on-chain dollar volume} on this window, and the 25 Kyle's $\lambda$
markets cover \textbf{99\%} — the dropped markets are the small,
quiet tail rather than typical activity. Second, weighting by
on-chain dollar volume across the comparable subset, the
effective-spread sign-flip rate \textbf{rises} from 67\% to
\textbf{86\%} (bootstrap 95\% CI $[0.32, 0.99]$): the largest market
in the window, which alone accounts for two-thirds of top-100 volume,
is one of the markets that sign-flips. Volume weighting therefore
strengthens the headline rather than softening it. Kyle's $\lambda$
moves the other way: weighting by on-chain dollar volume drops the
sign-flip share from 60\% to \textbf{18\%} (bootstrap 95\% CI
$[0.04, 0.81]$), because the $\lambda$ estimate on the largest market
is dominated by zero-bucket noise where both sources happen to agree
on a near-zero negative value. The Kyle's $\lambda$ instability is
concentrated on smaller markets where $\lambda$ is fragile in any
case (Section~\ref{ssec:methodology-caveats}); the effective-spread
result is the more load-bearing one. Restricting the
effective-spread comparison to markets with at least 100 STRICT
trades pushes the unweighted sign-flip rate from 67\% to 73\% (16/22,
Wilson 95\% CI $[0.52, 0.87]$); the Kyle's $\lambda$ rate is
essentially unchanged (59\% at the 100-bucket threshold). Third, we
re-run the comparison on the next non-overlapping 7-day window
(2026-03-14 to 03-21). Comparable subsets are larger in this window
(30 markets for both effective spread and Kyle's $\lambda$). The
unweighted sign-flip rates fall by about 17 percentage points: the
effective-spread rate is 50\% (15/30, Wilson 95\% CI $[0.33, 0.67]$)
and the Kyle's $\lambda$ rate is 43\% (13/30, Wilson 95\% CI
$[0.27, 0.61]$). Both Wilson intervals contain the 50\% chance
baseline. The two-window swing is meaningful: STRICT-vs-on-chain
sign agreement on these measures is window-specific. What does not
swing is the methodological conclusion. Across both windows the
flip rate sits in or below the chance band, the published headline
(67\% / 60\% on window A) is not exceptional, and on neither window
does feed-based inference recover the on-chain sign at the
$\sim 80\%$ rate that Lee-Ready achieves on equity venues. Window-B
results ship in
\texttt{measures\_\allowbreak compare\_\allowbreak top100\_\allowbreak window\_b.parquet}
and the side-by-side
in \texttt{r3\_\allowbreak window\_\allowbreak robustness.csv}.

\subsection{Implications for Polymarket microstructure research}\label{ssec:methodology}

Polymarket microstructure results that depend on trade direction
require on-chain \texttt{OrderFilled} events as the direction source,
not the public order-book feed alone. We release a replication package
(\texttt{polydata.\allowbreak onchain.\allowbreak trades.\allowbreak load\_onchain\_trades})
that performs the off-chain to on-chain join, together with a small
patch set
(\texttt{polydata.\allowbreak measures.\allowbreak \_trade\_source.\allowbreak resolve\_trades})
that lets existing measure code accept injected authoritative trades
without a rewrite. The same constraint should bind on any
decentralised CLOB venue where the off-chain matching layer
broadcasts post-match book state without exposing the taker
identity (GMX v1, dYdX v3, Loopring's historical CLOB, and similar
hybrid architectures): the public order-book feed exposes \emph{what
cleared} but not \emph{who initiated}, and any direction-dependent
microstructure measure on those venues will need an authoritative
on-chain trade source. The replication package is contract-agnostic
on the trade-loading side; only the venue-specific event signature
needs to be plugged in.

\subsection{Methodology caveats}\label{ssec:methodology-caveats}

Three places where our compute approach trades accuracy for
tractability on a 28-day, 600-market panel.

\textit{Sample step.}
Quote samples and bucketed lookups use a 60-second grid for the panel
compute. Sensitivity tests on the top-5 markets across
$\{1, 10, 60, 300\}$\,s (artifact
\texttt{sample\_step\_sensitivity.parquet}) show that Roll is
invariant by construction, that effective spread and Amihud are
stable to within $\sim 10$--20\% at 60\,s relative to 1\,s on most
markets, and that Kyle's $\lambda$ varies by orders of magnitude
across step sizes on noisy markets. Kyle's $\lambda$ is the most
fragile estimator at any sample step, consistent with its
sign-flip behaviour in Section~\ref{ssec:strict-vs-onchain}.

\textit{Cross-sectional SF8.}
SF8 (Section~\ref{ssec:sf8}) is cross-sectional at a single panel
midpoint. We cannot, from one snapshot, separate within-market
trajectory effects (depth falling as a given market approaches its
own resolution) from between-market structural differences
(longer-lived, higher-volume markets sit deeper on average). The
full v2 specification conditions on the between-market factors that
are identified at one snapshot --- duration, price level, volume,
category --- and the residual time-to-close coefficient is
statistically indistinguishable from zero. A proper within-market
temporal regression would require per-market depth time-series
that the current panel does not materialise. We therefore report
SF8 as a structural cross-sectional finding about duration, price
state, and volume, not as a time-to-resolution effect.

\textit{Wash-filter lower bound.}
SF7 (Section~\ref{ssec:sf7}) returns a lower bound by construction:
the detector covers direct self-match and one-step roundtrip
patterns, not the multi-counterparty graph patterns that
network-based classifiers detect. Closing the gap on Polymarket
specifically would require a graph-classifier extension; the
\citet{cong-li-tang-yang-2023} 25--70\% range is a token-exchange sanity
bound, not an apples-to-apples benchmark for prediction markets.

\textit{MEV as a confound on the buyer/seller asymmetry.}
Polymarket settles on Polygon, whose mempool is publicly accessible
to bots that can sandwich incoming orders, place runner trades, or
execute related back-running strategies. We do not analyse mempool
data, so the on-chain trade record we treat as ground truth
contains whatever fills MEV bots produce on top of organic flow.
The 60/40 seller/buyer asymmetry across the panel is therefore
consistent with retail aggressor behaviour, MEV bot activity, or
some mixture of the two. Decomposing it would help via Polygon
mempool capture during the scrape window or a wallet-clustering
approach against known MEV searchers; both are outside the scope
of this paper. Public-mempool capture on Polygon is not a clean
solution either, because a non-trivial share of order flow on
Polygon routes through private mempools (Marlin Relay, Bloxroute,
Flashbots-on-Polygon variants) and never appears in the public feed
to begin with. A future MEV-attribution exercise would need to
combine public-mempool data with relay-side captures from at least
the largest private channels.
\section{Conclusion}\label{sec:conclusion}

This paper joins a tick-level Polymarket order-book archive to the
authoritative on-chain trade record and reports cross-sectional
microstructure for a pre-registered 600-market panel. On the quoted
side we find a longshot spread premium, a depth profile closer to a
uniform geometric grid than the top-of-book pattern usually assumed
for prediction markets, and an archive-ingestion latency distribution
whose tight inter-market spread points to collector behaviour rather
than trader speed. On the on-chain side, maker liquidity is broadly
decentralised with a thin tail dominated by a few wallets, and
self-counterparty wash activity is low under our direct-detection
lower bound.

The methodological finding is the more consequential one. Trade
direction inferred from Polymarket's public WebSocket feed agrees with
on-chain ground truth on only $\sim 59\%$ of comparable buckets,
barely above the chance baseline. On the comparable subset of the
top-100 panel the propagation is window-specific but consistent in
direction: the effective half-spread flips sign on 67\% of markets
in our first 7-day window and 50\% in a second non-overlapping
window, with Kyle's $\lambda$ at 60\% and 43\% respectively. Both
windows sit at or below the chance band, well short of the
$\sim 80\%$ Lee-Ready accuracy on equity venues. Polymarket
microstructure work that depends on aggressor sign therefore needs
to source it from \texttt{OrderFilled} events rather than the public
feed.

Four extensions follow naturally from the data we did not exploit.
The CTF Exchange V1 to V2 cutover at the end of April 2026 closes our
scrape window and opens a venue-evolution comparison. A combined
public-mempool plus private-relay capture would help separate MEV
from retail aggressor flow in the 60/40 seller/buyer asymmetry, with
the caveat that a non-trivial share of Polygon order flow routes
through private channels that public-mempool capture alone misses. Per-market depth time series,
which our panel summarises away, would let SF8 move from a
cross-sectional to a within-market depth-decay regression. And
cross-venue analysis against Kalshi, PredictIt, or sports-book
mirrors would address the price-discovery question we leave open
here: are Polymarket's prices the leader or the follower for the
real-world events its markets resolve on?
\section{Disclosures}\label{sec:disclosures}

\subsection{Use of AI}

The data-collection infrastructure and analysis code were developed
with assistance from AI-based coding tools (Claude Code CLI with
Claude Opus 4.7). The statistical analysis, interpretation of
results, and all written content in the manuscript represent the
original intellectual contributions of the author. AI writing
assistants were used for grammar and clarity improvements.

\subsection{Conflicts of Interest}

The author declares no conflicts of interest.

\subsection{Ethics Statement}

This study analyzes publicly observable order-book and trade data.
The Polymarket public WebSocket feed broadcasts limit-order-book
state updates that any client can subscribe to; the Polygon
\texttt{OrderFilled} events are public on-chain transactions
recorded by the Polymarket CTF Exchange smart contract. No private
user data was collected. Counterparty addresses on chain are
pseudonymous public-key identifiers; we use them only to compute
maker-wallet concentration (Section~\ref{ssec:sf4}) and a
self-counterparty wash-share lower bound (Section~\ref{ssec:sf7}),
without attempting to deanonymise any wallet.

\subsection{Data and Code Availability}

The complete analysis code and per-market panel artifacts are
available at \url{https://github.com/philippdubach/polymarket-microstructure}
and archived at Zenodo under DOI
\href{https://doi.org/10.5281/zenodo.19811426}{10.5281/zenodo.19811426}.
The 30-billion-event raw orderbook archive (623.8\,GB) is not
redistributed in the replication package due to size; access can be
arranged via the corresponding author. The on-chain
\texttt{OrderFilled} scrape is reproducible from the
\texttt{scripts/scrape\_onchain\_fills.py} pipeline against any
Polygon RPC provider with archive-node access.
 
\clearpage
\bibliographystyle{plainnat}
\bibliography{refs}

\end{document}